\documentclass[a4paper,fleqn,amssymb,usenatbib,useAMS,usedcolumns]{mnras}
\usepackage{times} 
\usepackage{latexsym,amssymb,amsmath,amsfonts}
\usepackage{rotfloat} 
\usepackage{rotating} 
\usepackage{float}
\usepackage{graphicx}
\usepackage{natbib}
\usepackage{supertabular}
\usepackage{longtable}
\usepackage{url}
\usepackage{dcolumn}
\usepackage{placeins}
\usepackage{multirow} 
\usepackage{textcomp}
\usepackage{multicol}
\usepackage{bm}
\usepackage{pdflscape}
\usepackage[T1]{fontenc}
\usepackage[caption=false]{subfig}
\begin{document}
\newcommand{\hi}{\mbox{H\,{\sc i}}}
\newcommand{\mgii}{\mbox{Mg\,{\sc ii}}}
\newcommand{\mgi}{\mbox{Mg\,{\sc i}}}
\newcommand{\feii}{\mbox{Fe\,{\sc ii}}}
\newcommand{\caii}{\mbox{Ca\,{\sc ii}}}
\newcommand{\nai}{\mbox{Na\,{\sc i}}}
\def\h2{$\rm H_2$}
\def\Nh2{$N$(H${_2}$)}
\def\chin{$\chi^2_{\nu}$}
\def\chiu{$\chi_{\rm UV}$}
\def\lya{\ensuremath{{\rm Ly}\alpha}}
\def\lymana{\ensuremath{{\rm Lyman}-\alpha}}
\def\kms{km\,s$^{-1}$}
\def\cms{cm$^{-2}$}
\def\cc{cm$^{-3}$}
\def\zabs{$z_{\rm abs}$}
\def\zem{$z_{\rm em}$}
\def\nhi{$N$($\hi$)}
\def\ln{log~$N$}
\def\nh{$n_{\rm H}$}
\def\ne{$n_{\rm e}$}
\def\21{21-cm}
\def\ts{$T_{\rm s}$}
\def\th{$T_{\rm 01}$}
\def\t0{$\tau_{\rm 0}$}
\def\ll{$\lambda\lambda$}
\def\l{$\lambda$}
\def\fc{$C_{\rm f}$}
\def\c21{$C_{\rm 21}$}
\def\mjb{mJy~beam$^{-1}$}
\def\taudv{$\int\tau dv$}
\def\ha{H\,$\alpha$}
\def\taudvl{$\int\tau dv^{3\sigma}_{10}$}
\def\taudv{$\int\tau dv$}
\def\wmg{$W_{\mgii}$}
\def\wfe{$W_{\feii}$}
\def\dgi{$\Delta (g-i)$}
\def\ebv{$E(B-V)$}
%
%=============================================================================================
%
\title[\hi\ \21\ absorption from strong \mgii\ absorbers]{
{\hi\ \21\ absorption from $z\sim0.35$ strong \mgii\ absorbers
}
\author[R. Dutta et al.]{R. Dutta$^1$\thanks{E-mail: rdutta@iucaa.in}, R. Srianand$^1$\thanks{E-mail: anand@iucaa.in}, N. Gupta$^1$\thanks{E-mail: ngupta@iucaa.in}, R. Joshi$^1$\thanks{E-mail: rjoshi@iucaa.in} \\ 
$^1$ Inter-University Centre for Astronomy and Astrophysics, Post Bag 4, Ganeshkhind, Pune 411007, India \\
} 
}
\date{Accepted. Received; in original form }
\pubyear{}
\maketitle
\label{firstpage}
\pagerange{\pageref{firstpage}--\pageref{lastpage}}
%
%============================== ABSTRACT =================================================================================
%
\begin {abstract}  
\par\noindent
We have searched for \hi\ \21\ absorption in 11 strong \mgii\ systems ($W_{\rm r}$(\mgii\ $\lambda$2796) $\ge$ 1 \AA) at $0.3<z<0.5$ using the Giant Metrewave Radio Telescope. 
We have detected \hi\ \21\ absorption in two of these. From the integrated optical depth (\taudv) we estimate \nhi\ = 43 $\pm$ 2 and 9 $\pm$ 2 in units of $10^{19}$\,\cms\ 
for the absorbers towards J1428$+$2103 (\zabs\ = 0.3940) and J1551$+$0713 (\zabs\ = 0.3289), respectively, assuming spin temperature, \ts\ = 100 K, and gas covering factor, 
\fc\ = 1. The velocity width of the \hi\ absorption towards J1428$+$2103 and J1551$+$0713 indicate that the gas temperature is $<$1600 K and $<$350 K, respectively. The $3\sigma$ 
upper limits on \taudv\ in case of the \hi\ \21\ non-detections indicate that these \mgii\ absorbers are likely to arise from sub-damped Lyman-$\alpha$ systems, when we assume 
\ts\ = 100 K and \fc\ = 1. This is verified for one of the systems which has \nhi\ measurement using Lyman-$\alpha$ absorption detected in the ultraviolet spectrum. We estimate 
the detection rate of \hi\ \21\ absorption in strong \mgii\ systems in our sample to be 0.18$^{+0.24}_{-0.12}$ at $z\sim0.35$, for an integrated optical depth sensitivity of 
$\le$ 0.3 \kms. Comparing with the results of \hi\ \21\ absorption surveys in strong \mgii\ systems at higher redshifts from the literature, we do not find any significant 
evolution in the incidence and number density per unit redshift of \hi\ \21\ absorbers in strong \mgii\ systems over $0.3<z<1.5$.
\end {abstract}  
%
%=========================== KEY WORDS ===================================================================================== 
%
%
%=========================== INTRODUCTION ================================================================================== 
%
\section{Introduction} 
\label{sec_introduction}  
\hi\ \21\ absorption is an excellent tool to study the physical conditions of high column density cold ($\sim$100 K) neutral gas in galaxies \citep{kulkarni1988}. 
Studies of \hi\ \21\ absorption at different redshifts can be used to trace the redshift evolution of the cold gas fraction in galaxies, and hence understand the 
processes driving the observed redshift evolution of the global star formation rate density \citep[SFRD;][]{madau2014}. There have been several \hi\ \21\ absorption 
searches towards radio-loud quasars that show intervening absorption in their optical or ultraviolet (UV) spectra 
\citep[e.g.][]{briggs1983,lane2000,kanekar2003,curran2005,gupta2009,kanekar2009,curran2010,srianand2012,gupta2012,kanekar2014,dutta2017b},
or are at small impact parameters ($b$) from foreground galaxies 
\citep[e.g.][]{carilli1992,gupta2010,gupta2013,borthakur2011,borthakur2016,zwaan2015,reeves2015,reeves2016,dutta2016,dutta2017a}.
These studies have indicated that the distribution of cold \hi\ gas around low-$z$ ($z<$ 0.4) galaxies is patchy \citep[see][]{srianand2013},
with an average covering factor of $\sim$20\% within $b\sim$35 kpc \citep[see][]{dutta2017a}, and that the \hi\ gas at high-$z$ ($z>$ 2) 
is predominantly warmer (spin temperature, \ts\ $\gtrsim$ 1000 K) compared to the Milky Way \citep[see][]{srianand2012,kanekar2014}. However, 
careful mapping and interpretation of the redshift evolution of the cold gas fraction based on the present measurements is difficult due to 
different sample selection techniques used at different redshifts, and the small number of \hi\ \21\ detections known till date.

Strong \mgii\ absorbers [i.e. rest equivalent width (REW) of \mgii\ $\lambda 2796$, \wmg\ $\ge$ 1.0 \AA] have been shown to be associated with galaxies 
\citep{bergeron1986,lanzetta1990,bergeron1991,steidel1995,guillemin1997,churchill2005,chen2010a,rao2011,bowen2011}, and trace gas with high \nhi\ \citep{rao2006},
like damped Lyman-$\alpha$ absorbers (DLAs) and sub-DLAs which have \nhi\ $\ge$ 2$\times$10$^{20}$\,cm$^{-2}$ and $\sim$ 10$^{19}$ $-$ 2$\times$10$^{20}$\,cm$^{-2}$, 
respectively \citep[see for a review][]{wolfe2005}. Hence, they are appropriate targets to carry out \hi\ \21\ absorption searches, especially at $z<$ 1.5 
where ground-based observations of DLAs are limited by the atmospheric cutoff of light below 3000 \AA. Large homogeneous samples of \mgii\ absorbers are available 
\citep[e.g.][]{nestor2005,prochter2006,quider2011,zhu2013}, thanks to the large database of quasar spectra in the Sloan Digital Sky Survey \citep[SDSS;][]{york2000}. 
Taking advantage of this, there have been systematic searches for \hi\ \21\ absorption in SDSS-selected samples of \mgii\ systems at 0.5 $<z<$ 1.5 
\citep{gupta2007,gupta2009,gupta2012,kanekar2009,dutta2017b}. \citet{gupta2012} have shown that the detection rate of \hi\ \21\ absorption in strong \mgii\ 
systems remains constant over this redshift range, within the uncertainties, which is intriguing given the proposed connection between redshift evolution of 
the global SFRD and strong \mgii\ absorption \citep[e.g.][]{menard2011,chen2016}. \citet{dutta2017b} have shown that strong \mgii\ absorbers have higher 
probability being DLAs and giving rise to \hi\ \21\ absorption when strong \feii\ absorption (REW of \feii\ $\lambda$2600, $W_{\feii}$ $\ge$ 1 \AA) is present. 
However, even the strong \feii\ systems show no clear evolution in incidence of \hi\ \21\ absorption at 0.5 $<z<$ 1.5.

On the other hand, the incidence of \hi\ \21\ absorption in DLAs increases from $\sim$20\% at $z>2$ to $\sim$60\% at $z<1$ 
\citep{kanekar2009,srianand2012,dutta2017b}. Further, the incidence of \h2\ absorption in DLAs/sub-DLAs increases from $\sim$10\% 
at $z\ge1.8$ \citep{noterdaeme2008} to $\sim$50\% at $z\le0.7$, with most of the \h2\ absorption at low-$z$ arising from sub-DLAs
at $b>$ 10 kpc from the host galaxies \citep{muzahid2015}. Hence, the physical conditions and origin of cold gas around galaxies 
could be significantly different at low-$z$. The lower redshift cutoff of the \mgii\ samples searched for \hi\ \21\ absorption has 
usually been chosen as $z\sim$ 0.5 due to the difficulty of identifying large number of \mgii\ absorbers in the low signal-to-noise 
ratio (SNR) blue part of the SDSS spectra. Till date there have been only few searches for \hi\ \21\ absorption in strong \mgii\ 
selected systems at $z<$ 0.5 \citep[see][and references therein]{lane2000}. Now due to the increased wavelength coverage and SNR 
of SDSS-Baryon Oscillation Spectroscopic Survey (BOSS) in the blue part, \mgii\ absorption can be searched upto $z$ = 0.3 in the 
quasar spectra.

Here we wish to extend the \hi\ \21\ absorption studies in strong \mgii\ systems to lower redshifts (i.e. $z<$ 0.5), in order to study 
the redshift evolution of incidence of \hi\ \21\ absorption in these systems. Additionally, it is relatively easier to identify and study 
the properties of host galaxies at these redshifts. Hence, this will allow us to connect between \hi\ \21\ studies in 0.5 $<z<$ 1.5 
strong \mgii\ systems and $z<$ 0.4 galaxy-selected samples. Further, due to the lack of a statistically significant DLA sample at low-$z$, 
the evolution of the cosmic \hi\ mass density ($\Omega_{\hi}$) at $z<$ 1.5 has been a matter of debate \citep[e.g.][]{rao2006,neeleman2016}. 
Hence, studying how the incidence of cold \hi\ gas evolves over this redshift range will help to shed light on the redshift evolution of $\Omega_{\hi}$.

We present here the results from our Giant Metrewave Radio Telescope (GMRT) search for \hi\ \21\ absorption in eleven strong \mgii\ 
systems at 0.3 $<z<$ 0.5 selected from SDSS. Our sample has increased the number of observed strong \mgii\ systems in this redshift 
range \citep[see][for the previous compilation, and Section~\ref{sec_discussion1} for further details]{lane2000} by a factor of three. 
This paper is structured as follows. We describe our sample in Section \ref{sec_sample} and our GMRT observations in Section \ref{sec_observations}. 
We present the results from our search for \hi\ \21\ absorption in Section \ref{sec_results}. We discuss and summarize our results in Sections \ref{sec_discussion} 
and \ref{sec_summary}, respectively. Throughout this work we use a flat $\Lambda$-cold dark matter cosmology with $H_{\rm 0}$ = 70\,\kms~Mpc$^{-1}$ and $\Omega_{\rm M}$ = 0.30.
%
%==================================================== OBSERVATIONS =========================================================
%
\section{Sample selection and properties}
\label{sec_sample}
We identified 298 quasars from SDSS Data Release 12 \citep[DR12;][]{alam2015} with: (a) peak flux density at 1.4 GHz
$>$ 50 mJy from the Faint Images of the Radio Sky at Twenty-Centimeters \citep[FIRST;][]{white1997} survey, to allow
a sensitive search for \hi\ \21\ absorption; and (b) \zem\ $<$1.95, such that the Lyman-$\alpha$ forest is below 3600\,\AA\
(i.e. not covered by the SDSS spectra). In the SDSS spectra of these quasars, we visually searched for strong \mgii\ 
absorption (\wmg\ $\ge$ 1.0 \AA) at 0.29$<$ \zabs\ $<$0.42. We identified 12 such \mgii\ absorbers, which form the 
complete sample of strong \mgii-selected systems at these redshifts that can be searched for \hi\ \21\ absorption 
using L-band of GMRT. The details of these absorbers are presented in Table~\ref{tab:sample}.

The median values of \wmg\ and the \mgii\ doublet ratio of the absorbers in our sample are 1.27 \AA\ and 1.25, respectively. 
\mgi\ absorption is detected in 8 of these strong \mgii\ systems (though the \mgi\ absorption towards J1428$+$2103 could be 
blended with a metal line from another absorber), with a median $W_{\rm r}$(\mgi\ $\lambda$2852)$/$\wmg\ = 0.39. \feii\ absorption 
lines from the strong \mgii\ systems are not covered in the SDSS spectra, except for J1628$+$4734 \footnote{In case of J0209$-$0438 
and J1608$+$1029, the expected position of the \feii\ $\lambda$2600 line falls right at the edge of the SDSS spectra where it is not 
possible to measure the equivalent width.}. In case of J1628$+$4734, absorption from \feii\ $\lambda$2600 and $\lambda$2586 are not 
detected in the SDSS spectrum, and we estimate a $3\sigma$ upper limit of $W_{\feii}$ $\le$ 0.4 \AA. Note that the \mgii\ absorption 
towards J1628$+$4734 has a very broad profile and no other associated absorption is detected. The presence of \caii\ and \nai\ absorption 
usually indicates that the gas has high \nhi. \caii\ absorption is detected from the \mgii\ systems towards J1514$+$2813 (REW of \caii\ 
$\lambda$3934, $W_{\caii}$ =  0.48 $\pm$ 0.09 \AA), J1551$+$0713 ($W_{\caii}$ = 0.30 $\pm$ 0.06 \AA), and J1619$+$3030 ($W_{\caii}$ = 
0.27 $\pm$ 0.07 \AA). \nai\ absorption is detected only from the \mgii\ system towards J1619$+$3030 (REW of \nai\ $\lambda$5891, 
$W_{\nai}$ = 0.11 $\pm$ 0.06 \AA).

In the SDSS images of all the quasars, we find at least one candidate host galaxy of the strong \mgii\ absorption at $b<$110 kpc, 
which has photometric redshift consistent with the absorber redshift within the uncertainties (see Table~\ref{tab:sample}). 
The SDSS $r$-band magnitude of the nearest host galaxy candidates are in the range of 20.2$-$22.2 with a median of 20.8, and their 
observed $g-r$ colours are in the range of 0.0$-$2.3 with a median of 1.1. The host galaxies could also be faint galaxies in proximity 
or superimposed to the bright quasars, and hence difficult to detect in the SDSS images. Nebular emission lines are usually detected
in such cases of `galaxy on top of quasars' \citep[e.g.][]{noterdaeme2010,york2012,straka2013}. However, we do not detect 
any nebular emission lines at the redshift of the \mgii\ absorption in the SDSS spectra of the quasars. The non-detection of 
[O\,{\sc ii}] $\lambda$3727 emission at the redshift of the \mgii\ absorption in the median stacked spectrum of the quasars
(after subtracting the quasar continuum) provides a $3\sigma$ upper limit on the average surface star formation rate density of 
4 $\times$ 10$^{-4}$ M$_\odot$\,yr$^{-1}$\,kpc$^{-2}$ \citep[following][]{kewley2004}. Deep images and spectroscopic redshifts 
of all the galaxies in the field around the quasars are required to understand the origin of the strong \mgii\ absorbers.

Recently, it has been shown that the systems with \hi\ \21\ absorption cause systematic reddening in the spectrum of the background 
quasar \citep{dutta2017b}. We estimate the reddening of the quasars in our sample due to the intervening \mgii\ absorption by fitting 
the quasar spectra using the SDSS composite quasar spectrum \citep{vandenberk2001}, reddened by the Milky Way, LMC and SMC extinction 
curves \citep{gordon2003}, following procedures outlined in \citet{srianand2008,srianand2013} and \citet{noterdaeme2009,noterdaeme2010}. 
We find that none of the quasars show any significant signatures of reddening, with \ebv\ ranging from $-$0.18 to 0.08 and a median \ebv\ 
of 0.02 (see Table~\ref{tab:sample}). Note that the negative \ebv\ values indicate that the quasar spectra are bluer than the SDSS composite 
spectrum. For comparison, the median \ebv\ of the quasars showing \hi\ \21\ absorption in strong \feii\ systems at 0.5 $<z<$ 1.5 is 0.10 \citep{dutta2017b}.

Very Long Baseline Array (VLBA) 2.3 GHz images are available for six of the radio sources in our sample from the VLBA Calibrator Survey 
(VCS)\footnote{http://www.vlba.nrao.edu/astro/calib/}. We estimate the covering factor of the absorbing gas, \fc, by assuming it to be 
the core fraction, i.e. assuming that the absorbing gas covers only the core component seen in the VLBA image. Hence, we estimate \fc\ 
from the ratio of the peak flux density in the VCS image to the total arcsecond-scale flux density at 2.3 GHz [obtained by interpolating 
flux densities available from the NASA extragalactic data base \footnote{https://ned.ipac.caltech.edu/} \citep{gregory1991,condon1998}].
The \fc\ values are provided in Table~\ref{tab:sample}.
\begin{table*} 
\caption[Sample of strong \mgii\ systems at $z\sim0.35$.]{Sample of strong \mgii\ systems at $z\sim0.35$ toward radio-loud quasars.}
\centering
\begin{tabular}{cccccccccc}
\hline
Quasar & SDSS & \zem\ & \zabs\ & $W_{\rm r}$(\mgii\ $\lambda$2796) & $W_{\rm r}$(\mgii\ $\lambda$2803) & $W_{\rm r}$(\mgi\ $\lambda$2852) & $b$   & \ebv\ & \fc\ \\
name   & name &       &        & (\AA)                             & (\AA)                             & (\AA)                            & (kpc) &       &      \\
(1)    & (2)  & (3)   & (4)    & (5)                               & (6)                               & (7)                              & (8)   & (9)   & (10) \\
\hline
J0200$+$0322 & J020040.81$+$032249.4 & 1.581 & 0.3574 & 1.37 $\pm$ 0.13 & 0.97 $\pm$ 0.10 & $\le$0.41           & 109     & $-$0.15 & 0.62 \\
J0209$-$0438 & J020930.77$-$043826.1 & 1.131 & 0.3903 & 1.12 $\pm$ 0.07 & 0.74 $\pm$ 0.07 & 0.40 $\pm$ 0.06     & 65      & $+$0.05 & 0.64 \\ 
J0255$+$0253 & J025509.76$+$025345.6 & 0.663 & 0.3375 & 1.32 $\pm$ 0.20 & 1.53 $\pm$ 0.15 & 0.52 $\pm$ 0.14     & 15      & $+$0.01 & ---  \\
J0859$+$1552 & J085943.79$+$155232.8 & 0.616 & 0.3798 & 1.22 $\pm$ 0.12 & 0.98 $\pm$ 0.12 & 0.38 $\pm$ 0.09     & 33      & $-$0.06 & ---  \\
J1423$+$5150 & J142329.98$+$515008.9 & 1.685 & 0.3052 & 1.14 $\pm$ 0.33 & 0.58 $\pm$ 0.33 & $\le$0.74           & 98      & $+$0.02 & ---  \\
J1428$+$2103 & J142846.41$+$210336.6 & 1.454 & 0.3940 & 1.56 $\pm$ 0.26 & 1.12 $\pm$ 0.23 & 1.29 $\pm$ 0.29$^a$ & 26      & $-$0.01 & 0.53 \\
J1501$+$5619 & J150124.63$+$561949.7 & 1.466 & 0.4011 & 1.13 $\pm$ 0.15 & 1.18 $\pm$ 0.19 & 0.46 $\pm$ 0.12     & 17      & $+$0.07 & ---  \\
J1514$+$2813 & J151402.50$+$281334.8 & 1.548 & 0.3715 & 1.57 $\pm$ 0.13 & 1.35 $\pm$ 0.13 & 0.39 $\pm$ 0.08     & 46      & $+$0.08 & ---  \\
J1551$+$0713 & J155121.13$+$071357.7 & 0.675 & 0.3289 & 1.00 $\pm$ 0.07 & 1.03 $\pm$ 0.06 & 0.41 $\pm$ 0.06     & 99$^b$  & $-$0.11 & ---  \\
J1608$+$1029 & J160846.20$+$102907.7 & 1.232 & 0.3725 & 1.27 $\pm$ 0.11 & 0.99 $\pm$ 0.10 & $\le$0.44           & 57      & $+$0.07 & 0.94 \\
J1619$+$3030 & J161902.49$+$303051.6 & 1.288 & 0.3024 & 1.21 $\pm$ 0.10 & 1.27 $\pm$ 0.10 & 0.32 $\pm$ 0.08     & 34      & $+$0.05 & 0.48 \\
J1628$+$4734 & J162837.50$+$473410.5 & 1.632 & 0.4184 & 1.48 $\pm$ 0.11 & 1.43 $\pm$ 0.11 & $\le$0.43           & 95      & $-$0.18 & 0.91 \\
\hline
\end{tabular}
\label{tab:sample}
\begin{flushleft} {\it Notes.}
Column 1: quasar name used throughout this work. Column 2: SDSS (J2000) name of the quasar. Column 3: redshift of quasar. Column 4: redshift of intervening \mgii\ system. 
Columns 5, 6 and 7: REWs (\AA) of the \mgii\ $\lambda$2796, \mgii\ $\lambda$2803 and \mgi\ $\lambda$2852 absorption lines, respectively, measured by us. 
Column 8: impact parameter (kpc) assuming that the galaxy nearest to the quasar, with consistent photometric redshift in the SDSS images, is the host galaxy.
Column 9: \ebv\ of the quasar from spectral energy distribution (SED) fitting, assuming that the reddening is due to the intervening \mgii\ absorption. 
Negative values indicate that the quasars are bluer than the template used. Typical systematic error in the SED-fitting method due to the dispersion of the unreddened quasar SED is $\sim$0.1 \citep{dutta2017b}.
Column 10: covering factor (assumed to be the core fraction) determined from 2.3 GHz VLBA image of the background radio source (see Section~\ref{sec_sample} for details). \\
$^a$ Possibly blended. \\ 
$^b$ Host galaxy could also be behind the foreground galaxy superimposed on the quasar in the SDSS image (see Section~\ref{sec_j1551+0713}). 
\end{flushleft}
\end{table*}
\section{Observations and data reduction}
\label{sec_observations}
The observations were carried out using the L-band receiver on GMRT, with the 4 MHz baseband bandwidth split into 512 channels 
(spectral resolution $\sim$2\,\kms\ per channel, velocity coverage $\sim$1000\,\kms). Data were acquired in two polarization 
products, XX and YY. Standard calibrators were regularly observed during the observations for flux density, bandpass, and phase 
calibrations. The pointing centre was at the quasar coordinates and the band was centred at the redshifted \hi\ \21\ line frequency. 
The details of the radio observations of the sources are given in Table~\ref{tab:obslog}. The observations of J1619$+$3030 were 
affected by strong RFI and we were not able to obtain any useful data for it. 

\begin{figure*}
\subfloat[J0209$-$0438]{ \includegraphics[width=0.33\textwidth, bb = 50 50 740 750, clip=true]{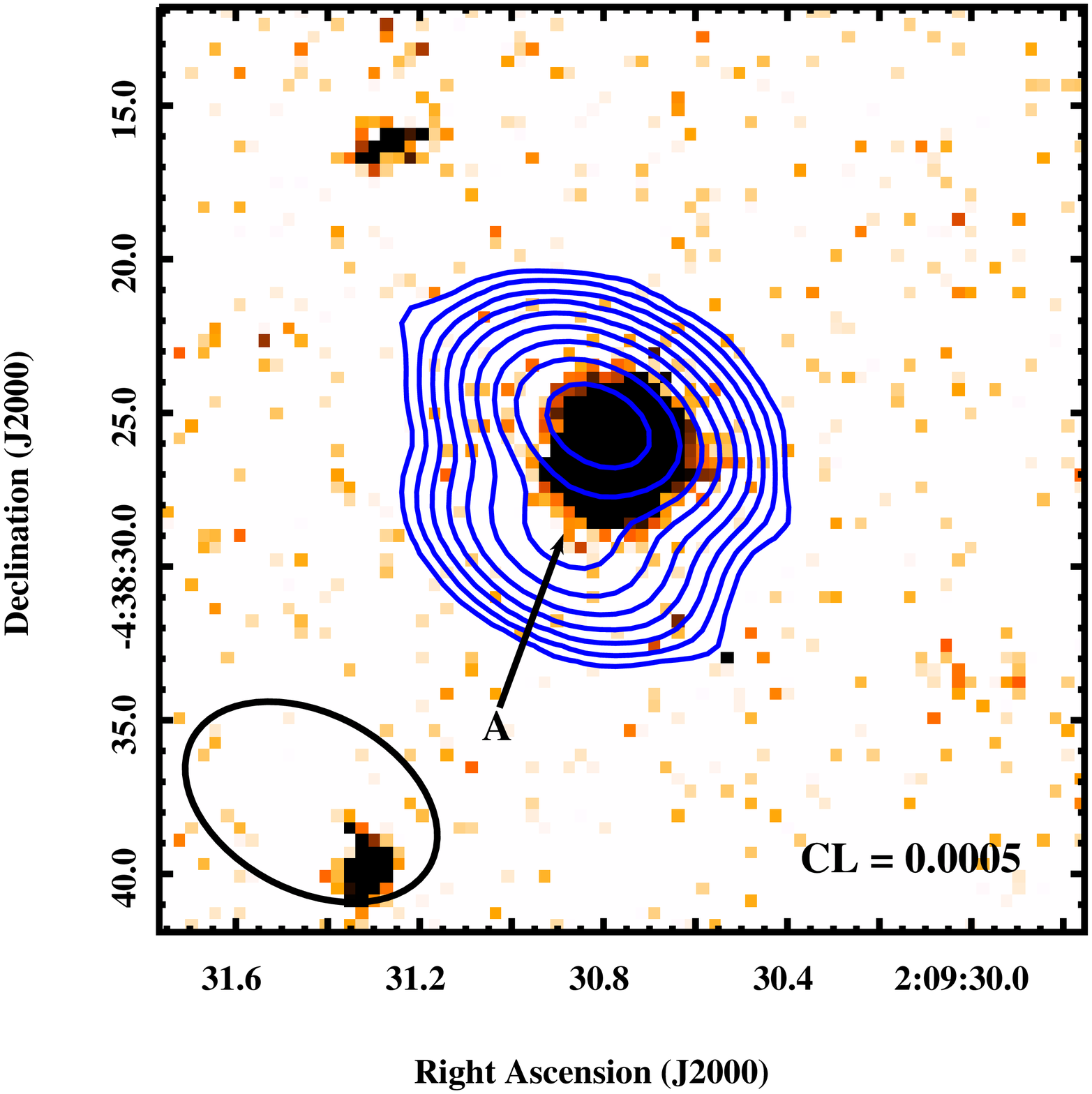} } 
\subfloat[J0255$+$0253]{ \includegraphics[width=0.34\textwidth, bb = 45 55 750 750, clip=true]{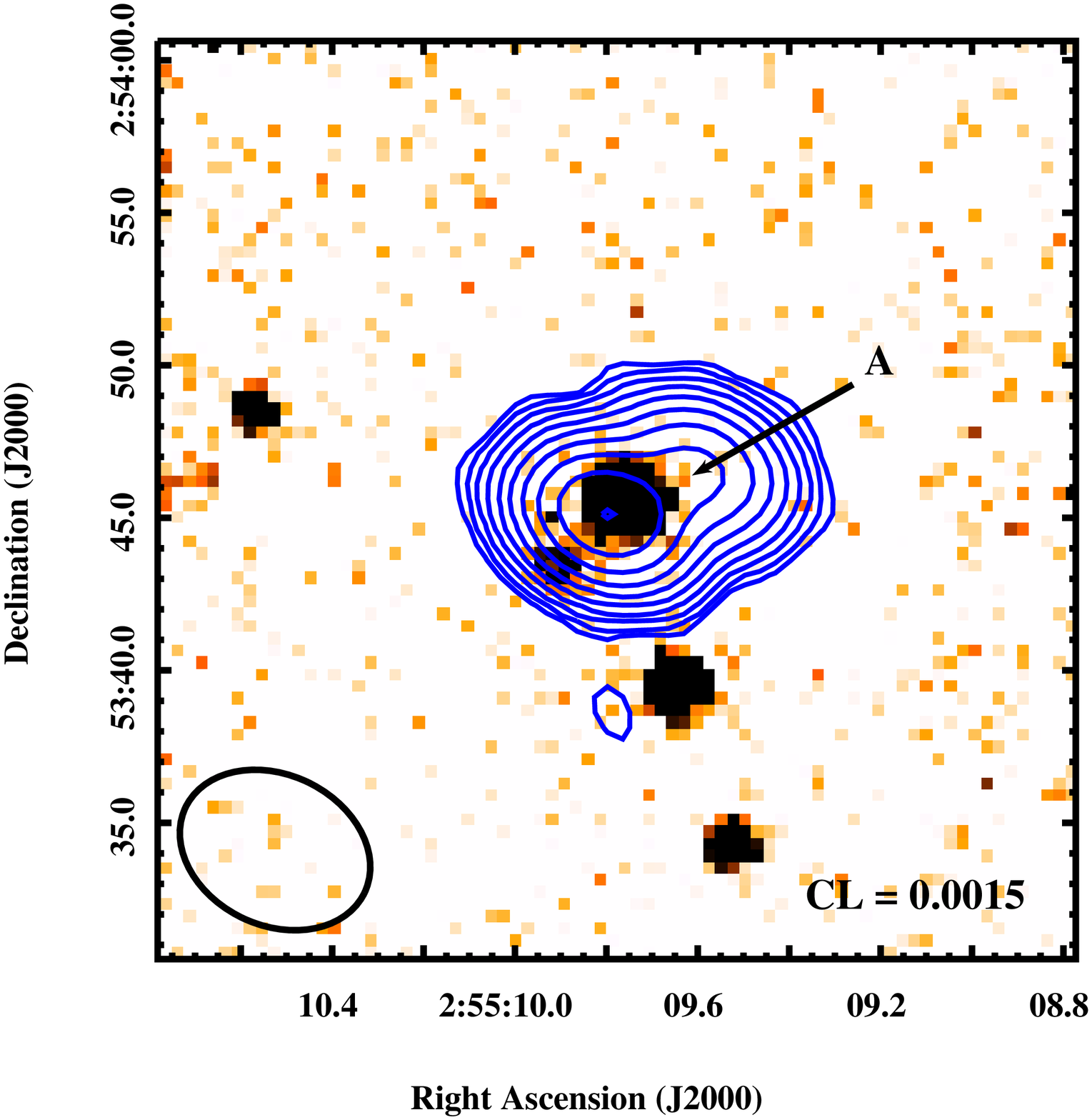} }
\subfloat[J1423$+$5150]{ \includegraphics[width=0.32\textwidth, bb = 40 30 730 730, clip=true]{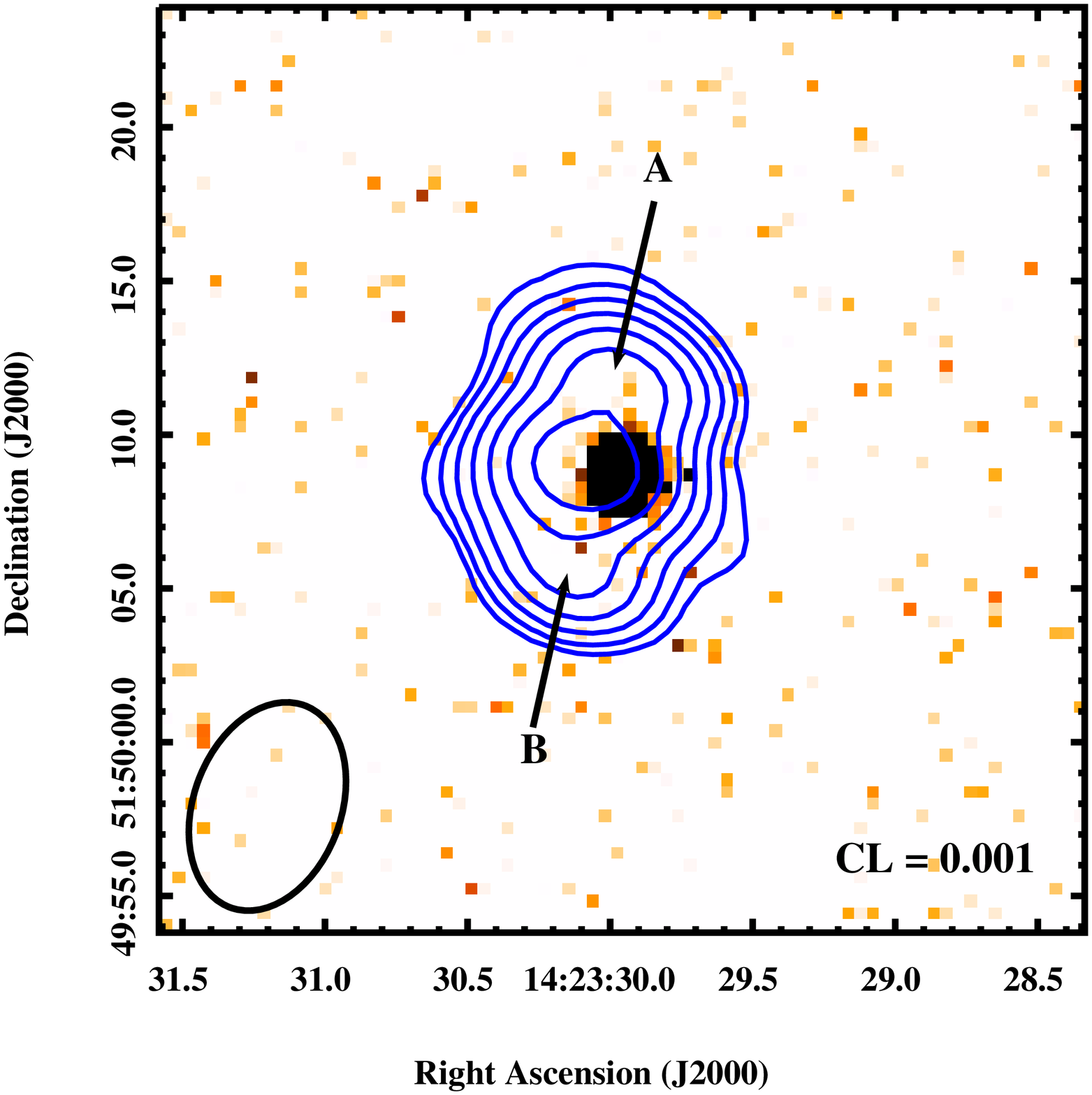} }
\caption{SDSS $r$-band images overlaid with the GMRT 1.4 GHz continuum contours of the radio sources: (a) J0209$-$0438, (b) J0255$+$0253 and (c) J1423$+$5150. 
The restoring beam of the continuum map is shown at the bottom left corner. The contour levels are plotted as CL $\times$ ($-$1,1,2,4,8,...)~Jy~beam$^{-1}$, 
where CL is given in the bottom right of each image. The rms in the radio images in (a), (b) and (c) are 0.0001, 0.0003 and 0.0002 Jy~beam$^{-1}$, respectively.
The continuum peaks of the lobes are indicated in each image.}
\label{fig:overlays} 
\end{figure*}

The data were reduced using the National Radio Astronomy Observatory (NRAO) Astronomical Image Processing System ({\sc aips}) 
following standard procedures as described in \citet{gupta2010}. Three of the radio sources (J0209$-$0438, J0255$+$0253 and 
J1423$+$5150) are extended in our GMRT continuum images (typical spatial resolution $\sim$2$''$), while the rest are compact. 
The radio continuum maps of these three extended sources overlaid on SDSS images are shown in Fig.~\ref{fig:overlays}.
The \hi\ \21\ absorption spectra were extracted at the locations of the continuum peak flux density in all cases, that also 
coincide well (within $\sim$1$''$) with the optical positions of the quasars in SDSS images. We also searched for \hi\ \21\ 
absorption towards the lobes in these three cases, but did not detect any significant absorption.
\begin{table} 
\caption[GMRT observation log of $z\sim0.35$ strong \mgii\ systems.]{GMRT observation log of $z\sim0.35$ strong \mgii\ systems.}
\centering
\begin{tabular}{cccc}
\hline
Quasar & Date & Time & $\delta v$ \\
       &      & (h)  & (\kms)     \\
(1)    & (2)  & (3)  & (4)        \\
\hline
J0200$+$0322 & 17 November 2015 & 5.7 & 2.4 \\
J0209$-$0438 & 8 December 2015  & 5.4 & 2.4 \\ 
J0255$+$0253 & 24 December 2015 & 0.9 & 2.4 \\
J0859$+$1552 & 13 November 2015 & 4.8 & 2.4 \\
             & 02 October 2016  & 5.2 & 2.4 \\
J1423$+$5150 & 14 November 2015 & 5.3 & 2.3 \\
J1428$+$2103 & 5 December 2015  & 5.1 & 2.4 \\
J1501$+$5619 & 8 December 2015  & 2.7 & 2.4 \\
             & 5 February 2016  & 2.6 & 2.4 \\
J1514$+$2813 & 27 December 2015 & 5.4 & 2.4 \\
J1551$+$0713 & 23 November 2014 & 5.3 & 2.3 \\
J1608$+$1029 & 9 December 2015  & 1.2 & 2.4 \\
J1619$+$3030 & 26 February 2016 & \multicolumn{2}{c}{---RFI---} \\
J1628$+$4734 & 26 December 2015 & 5.3 & 2.4 \\
             & 26 February 2016 & 5.0 & 2.4 \\
             & 04 August 2016   & 2.6 & 2.4 \\
             & 05 August 2016   & 3.8 & 2.4 \\
\hline
\end{tabular}
\label{tab:obslog}
\begin{flushleft} {\it Notes.}
Column 1: quasar name. Column 2: date of observation. Column 3: time on source in h. Column 4: channel width in \kms. \\
\end{flushleft}
\end{table}
%
%=================================================== RESULTS =============================================================== 
%
\section{Results}
\label{sec_results}
We have detected \hi\ \21\ absorption in two (at \zabs\ = 0.3940 towards J1428$+$2103, and at \zabs\ = 0.3289 towards J1551$+$0713) out of 11 strong 
\mgii\ systems. Tentative absorption features were seen towards J0859$+$1552 and J1628$+$4734, and hence these sources were reobserved. However, these 
tentative features are at less than 2$\sigma$ significance in the co-added spectra and they are not consistently produced in the individual polarization 
spectra. Therefore we consider these to be non-detections. The parameters estimated from the GMRT \hi\ \21\ absorption spectra are summarized in Table~\ref{tab:radiopara}. 
In case of the extended radio sources, results from spectra extracted towards the lobes, as marked in Fig.~\ref{fig:overlays}, are also provided. 
We list the standard deviation in the optical depth at $\sim$2\,\kms\ spectral resolution ($\sigma_\tau$), and the 3$\sigma$ upper limit on the 
integrated optical depth from spectra smoothed to 10\,\kms\ (\taudvl). In case of the detections, we provide the peak optical depth ($\tau_{\rm p}$), 
the total integrated optical depth (\taudv), the velocity width which contains 90\% of the total optical depth ($v_{\rm 90}$), and the velocity offset 
of the peak \hi\ \21\ optical depth from the strongest metal component in the SDSS spectrum ($v_{\rm off}$). Additionally, we estimate \nhi\ 
from \taudv\ of the detections, or 3$\sigma$ upper limit to it from \taudvl\ in case of non-detections, assuming \ts\ = 100 K, typical of the 
CNM in the Milky Way \citep{wolfire1995,heiles2003}, and covering factor, \fc\ = 1. We note that the small velocity widths ($v_{\rm 90}$ $\sim$ 
5 and 14\,\kms) of the two \hi\ \21\ absorption lines reported here support the correlation between $v_{\rm 90}$ of \hi\ \21\ absorption lines 
and redshift noted by \citet{dutta2017b}.

\begin{figure*}
\includegraphics[height=1.0\textwidth, angle=90]{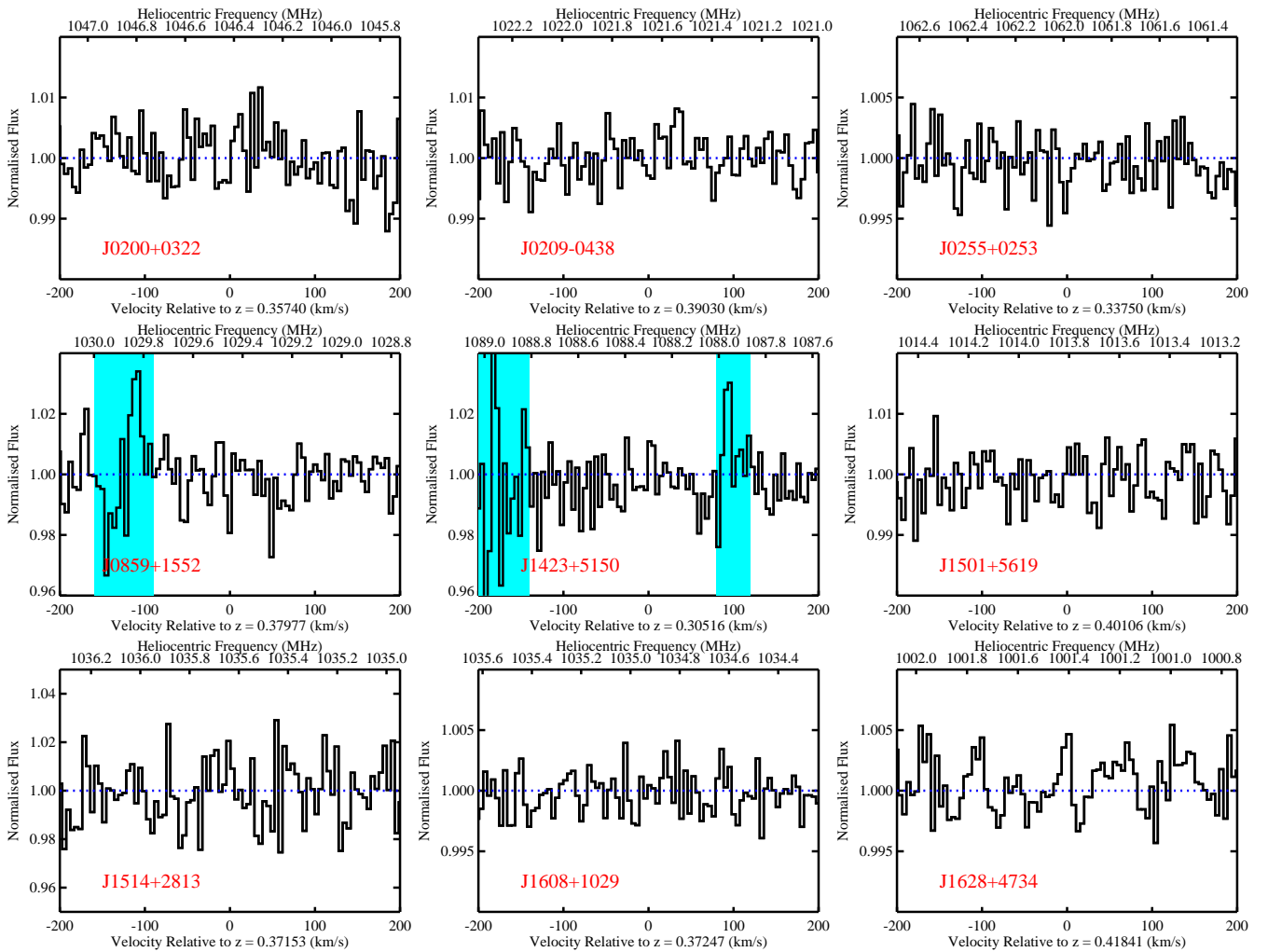}
\caption[GMRT \hi\ \21\ absorption spectra of the non-detections in the sample of strong \mgii\ systems.]
{\hi\ \21\ absorption spectra in case of the non-detections from the sample of strong \mgii\ systems at $z\sim0.35$ (smoothed to $\sim$5\,\kms\ for display purpose). 
The shaded regions mark frequency ranges affected by RFI. The quasar name as given in Table~\ref{tab:sample} is provided for each spectrum.}
\label{fig:21cmspectra}
\end{figure*}

\begin{figure*}
\subfloat[J1428$+$2103]{\includegraphics[height=0.5\textwidth, angle=90]{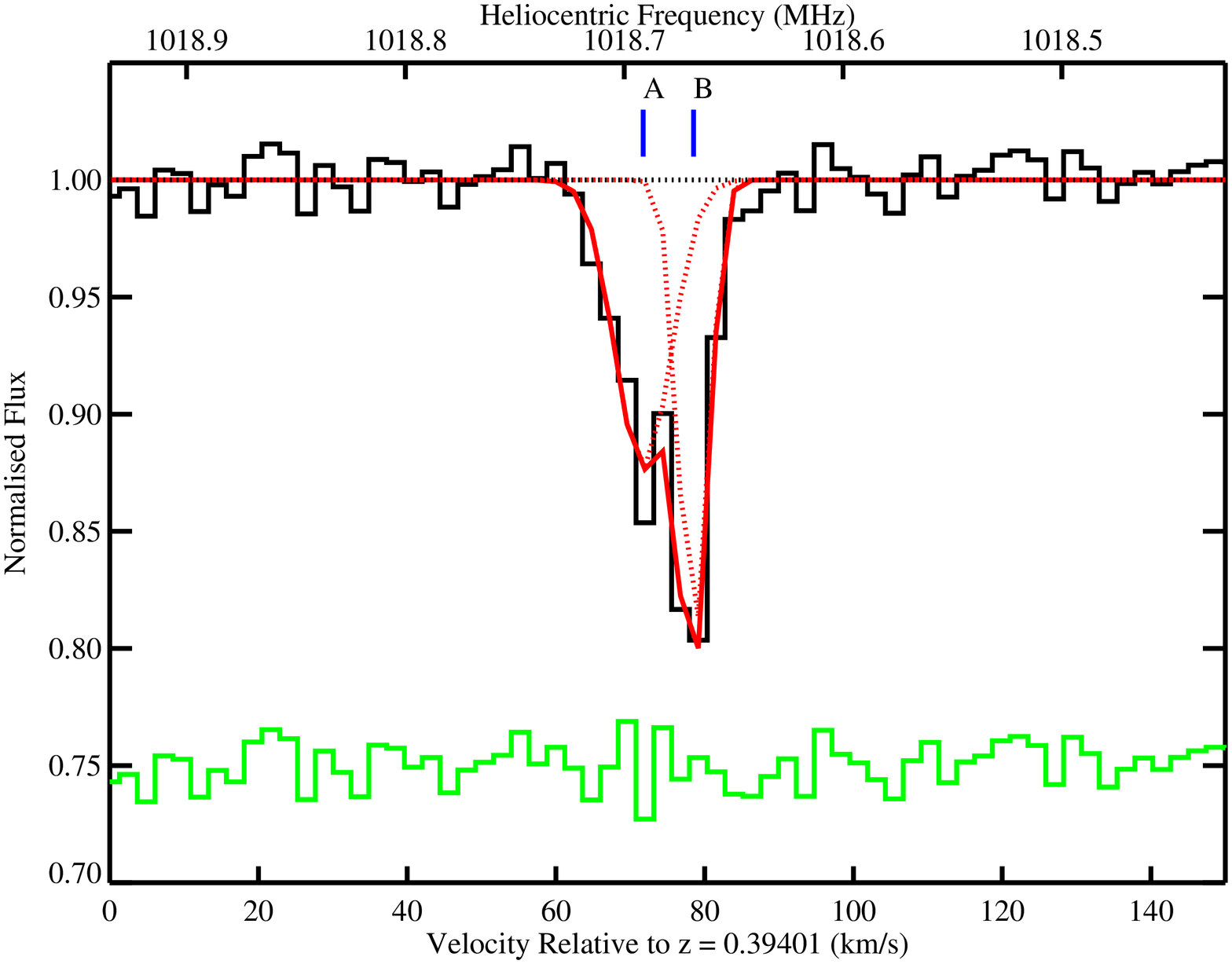}}
\subfloat[J1551$+$0713]{\includegraphics[height=0.5\textwidth, angle=90]{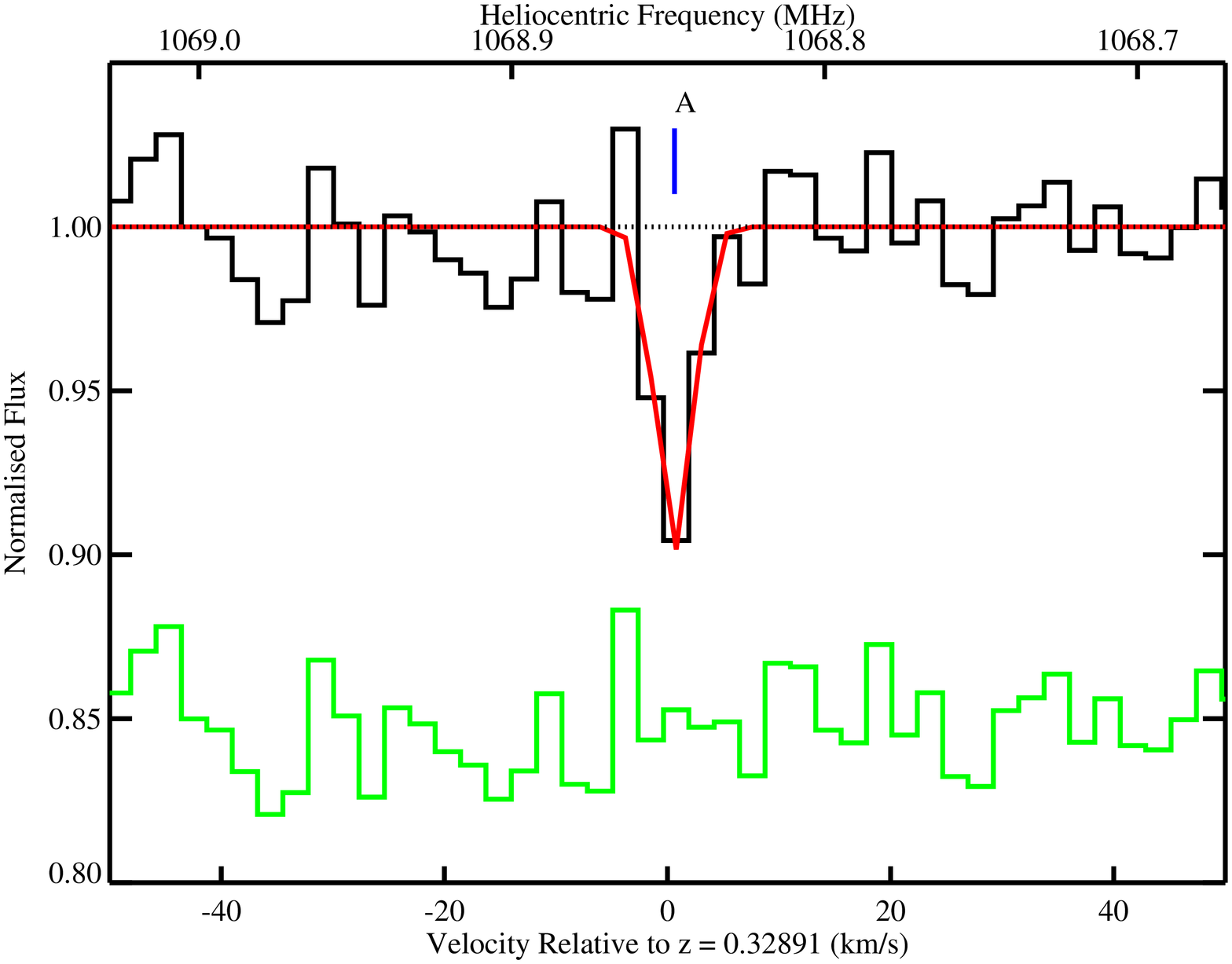}}
\caption[GMRT \hi\ \21\ absorption spectra of the detections in the sample of strong \mgii\ systems.]
{\hi\ \21\ absorption spectra in case of the detections: (a) J1428$+$2103 and (b) J1551$+$0713. 
Individual Gaussian components and the resultant fits to the absorption profiles are overplotted as dotted and continuous lines, respectively.
Residuals from the fit, shifted in the y-axis by an arbitrary offset for clarity, are also shown.
Locations of the peak optical depth of the individual components (as identified in Table~\ref{tab:gaussfit}) are marked by vertical ticks.}
\label{fig:21cmfit}
\end{figure*}

The \hi\ \21\ absorption spectra for the non-detections are shown in Fig.~\ref{fig:21cmspectra}. The Gaussian fits to the two \hi\ \21\ absorption 
are shown in Fig~\ref{fig:21cmfit}. The number of Gaussian components is determined based on the fit with \chin\ nearest to unity. Table~\ref{tab:gaussfit} 
gives details of the Gaussian fits, i.e. \zabs, full-width-at-half-maximum (FWHM) and $\tau_{\rm p}$ of individual Gaussian components. We also constrain 
the kinetic temperature, $T_{\rm k}$, and \nhi\ from the FWHM and $\tau_{\rm p}$ of the Gaussian components \citep[see section 4 of][]{dutta2017b}.
\begin{table*} 
\caption[Parameters derived from redshifted \hi\ \21\ absorption spectra of $z\sim0.35$ strong \mgii\ systems.]{Parameters derived from redshifted \hi\ \21\ absorption spectra of $z\sim0.35$ strong \mgii\ systems.}
\centering
\begin{tabular}{cccccccccc}
\hline
Quasar & Peak Flux & Spectral & $\sigma_\tau$ & $\tau_{\rm p}$ & \taudvl\ & \taudv\ & \nhi\                   & $v_{\rm 90}$ & $v_{\rm off}$ \\
       & Density   & rms      &               &                &          &         & (\ts$/100$ K) ($1/$\fc) &              &               \\
       & (\mjb)    & (\mjb)   &               &                & (\kms)   & (\kms)  & ($10^{19}$\,\cms)       & (\kms)       & (\kms)        \\
(1)    & (2)       & (3)      & (4)           & (5)            & (6)      & (7)     & (8)                     & (9)          & (10)          \\
\hline
J0200$+$0322    & 189 & 1.3 & $\le$0.007 & ---  & $\le$0.110 & ---             & $\le$2     & --- & --- \\
J0209$-$0438    & 202 & 1.0 & $\le$0.005 & ---  & $\le$0.096 & ---             & $\le$2     & --- & --- \\ 
J0209$-$0438    & 202 & 1.0 & $\le$0.005 & ---  & $\le$0.096 & ---             & $\le$2     & --- & --- \\ 
J0209$-$0438\_A &  45 & 1.1 & $\le$0.025 & ---  & $\le$0.337 & ---             & $\le$6     & --- & --- \\ 
J0255$+$0253    & 789 & 2.4 & $\le$0.003 & ---  & $\le$0.038 & ---             & $\le$1     & --- & --- \\
J0255$+$0253\_A & 271 & 2.3 & $\le$0.009 & ---  & $\le$0.122 & ---             & $\le$2     & --- & --- \\
J0859$+$1552    & 115 & 1.0 & $\le$0.009 & ---  & $\le$0.130 & ---             & $\le$2     & --- & --- \\
J1423$+$5150    & 120 & 1.3 & $\le$0.011 & ---  & $\le$0.107 & ---             & $\le$2     & --- & --- \\
J1423$+$5150\_A &  57 & 1.3 & $\le$0.022 & ---  & $\le$0.412 & ---             & $\le$8     & --- & --- \\
J1423$+$5150\_B &  32 & 1.3 & $\le$0.041 & ---  & $\le$0.735 & ---             & $\le$13    & --- & --- \\
J1428$+$2103    & 137 & 1.3 & $\le$0.010 & 0.22 & $\le$0.216 & 2.36 $\pm$ 0.08 & 43 $\pm$ 2 & 14  & 71  \\
J1501$+$5619    & 188 & 1.1 & $\le$0.006 & ---  & $\le$0.076 & ---             & $\le$1     & --- & --- \\
J1514$+$2813    & 57  & 1.0 & $\le$0.017 & ---  & $\le$0.289 & ---             & $\le$5     & --- & --- \\
J1551$+$0713    & 58  & 0.8 & $\le$0.014 & 0.10 & $\le$0.201 & 0.48 $\pm$ 0.09 & 9 $\pm$ 2  & 5   & 0   \\
J1608$+$1029    & 898 & 2.1 & $\le$0.002 & ---  & $\le$0.040 & ---             & $\le$1     & --- & --- \\
J1628$+$4734    & 330 & 0.8 & $\le$0.003 & ---  & $\le$0.060 & ---             & $\le$1     & --- & --- \\
\hline
\end{tabular}
\label{tab:radiopara}
\begin{flushleft} {\it Notes.}
Column 1: quasar name; in case of the extended radio sources, results from spectra extracted towards the lobes, as marked in Fig.~\ref{fig:overlays}, are also provided. 
Column 2: L-band peak flux density in \mjb. Column 3: spectral rms in \mjb. Column 4: \hi\ \21\ optical depth rms. 
Column 5: maximum \hi\ \21\ optical depth in case of \hi\ \21\ detections.
Column 6: 3$\sigma$ upper limit on the integrated \hi\ \21\ optical depth with data smoothed to 10 \kms. 
Column 7: integrated \hi\ \21\ optical depth in case of \hi\ \21\ detections.
Column 8: \nhi\ assuming \ts\ = 100 K and \fc\ = 1, in units of $10^{19}$\,\cms\ (3$\sigma$ upper limit in case of non-detections).
Column 9: velocity width which contains 90\% of the total optical depth in case of \hi\ \21\ detections. 
Column 10: velocity offset of the peak \hi\ \21\ optical depth from the strongest metal component in the SDSS spectrum in case of detections. \\
Note that the values given in Columns 3 and 4 are at the velocity resolution ($\sim$2\,\kms) specified in Column 4 of Table~\ref{tab:obslog}.
\end{flushleft}
\end{table*}
\begin{table*} 
\caption[Details of the Gaussian fits to the \hi\ \21\ absorption lines of $z\sim0.35$ strong \mgii\ systems.]{Details of the Gaussian fits to the \hi\ \21\ absorption lines.}
\centering
\begin{tabular}{ccccccc}
\hline
Quasar & ID  & \zabs\ & FWHM   & $\tau_{\rm p}$ & $T_{\rm k}$ & \nhi\                   \\
       &     &        &        &                &             & (\ts$/100$ K) ($1/$\fc) \\
       &     &        & (\kms) &                & (K)         & ($10^{19}$\,\cms)       \\
(1)    & (2) & (3)    & (4)    & (5)            & (6)         & (7)                     \\
\hline
J1428$+$2103 & A & 0.39434 & 8.6 $\pm$ 1.1 & 0.13 $\pm$ 0.03 & $<$1616 & 22 $\pm$ 9 \\
             & B & 0.39438 & 4.6 $\pm$ 0.4 & 0.22 $\pm$ 0.05 & $<$462  & 19 $\pm$ 6 \\
J1551$+$0713 & A & 0.32891 & 4.0 $\pm$ 0.1 & 0.10 $\pm$ 0.04 & $<$350  & 8 $\pm$ 4  \\         
\hline
\end{tabular}
\label{tab:gaussfit}
\begin{flushleft} {\it Notes.}
Column 1: quasar name. Column 2: absorption component as marked in Fig.~\ref{fig:21cmfit}. Column 3: redshift of absorption component. Column 4: FWHM (\kms) of Gaussian component. 
Column 5: peak optical depth of Gaussian component. Column 6: upper limit on $T_{\rm k}$ (K), obtained assuming the line width is purely due to thermal motions. 
Column 7: \nhi\ assuming \ts\ = 100 K and \fc\ = 1, in units of $10^{19}$\,\cms.
\end{flushleft}
\end{table*}
\subsection{\hi\ \21\ absorption towards J1428$+$2103}
\label{sec_j1428+2103}
The \hi\ \21\ absorption towards J1428$+$2103 is best fit with two Gaussian components, A and B, with FWHM $\sim$9\,\kms\ and $\sim$5\,\kms, respectively (see Table~\ref{tab:gaussfit}). 
The background radio source is compact in the 2.3 GHz VCS image. If we use the \fc\ obtained from this image (Table~\ref{tab:sample}), and assume \ts\ = 100 K, as in the Milky Way CNM, 
then we can constrain \nhi\ $\sim$ 8.1 $\times$ 10$^{20}$ \cms. On the other hand, using the constraint obtained on $T_{\rm k}$ from the FWHM, and assuming that the spin temperature 
follows the kinetic temperature \citep{field1959,bahcall1969,mckee1977,wolfire1995,liszt2001,roy2006}, we can obtain a conservative upper limit on the \nhi\ of the two absorption 
components A and B as, \nhi\ $<$ 6.7 $\times$ 10$^{21}$ \cms\ and $<$ 1.7 $\times$ 10$^{21}$ \cms, respectively. Hence, this absorption is likely to arise from a DLA. Note that 
the \hi\ associated with metal line absorption from different components of the gas need not necessarily be detected in \21\ absorption, either because it is too warm or it has 
low covering factor of the background radio source \citep[e.g.][]{srianand2012,rahmani2012,dutta2015}. Hence the total \nhi\ associated with the system could be larger.
No \caii\ and \nai\ absorption from this system is seen in the SDSS spectrum of the quasar. We estimate $3\sigma$ upper limit of $W_{\caii}$ $\le$ 0.5 \AA\ and $W_{\nai}$ $\le$ 0.5 \AA. 
We identify a possible host galaxy candidate located $\sim$5$''$ north of the quasar ($b\sim$ 26 kpc) in the SDSS image (Table~\ref{tab:sample}), which has a consistent photometric 
redshift ($z_{\rm phot}$ = 0.357 $\pm$ 0.089).
\subsection{\hi\ \21\ absorption towards J1551$+$0713} 
\label{sec_j1551+0713}
A narrow (FWHM $\sim$4\,\kms) \hi\ \21\ absorption is detected towards J1551$+$0713. No VLBA image of the background radio source is available. However, the radio source is compact in the 
$\sim$0.2 arcsec-scale 8.4 GHz image, with flux density of 30 mJy \citep{myers2003}. If we assume a \fc\ of unity and \ts\ $\approx$ $T_{\rm k}$ $<$ 350 K, then we can constrain the \nhi\ as $<$ 
3 $\times$ 10$^{20}$ \cms, whereas for \ts\ = 100 K, \nhi\ would be 9 $\times$ 10$^{19}$ \cms. Hence, the \hi\ \21\ absorption could arise from a sub-DLA. However, as noted in Section~\ref{sec_j1428+2103}, 
the total \nhi\ associated with the \mgii\ system could be larger. While \caii\ absorption from this system is detected in the SDSS spectrum of the quasar (see Section~\ref{sec_sample}), 
\nai\ absorption is not detected ($3\sigma$ upper limit of $W_{\nai}$ $\le$ 0.1 \AA). The measured $W_{\caii}$ is similar to that of the weak \caii\ absorber population found in SDSS spectra 
\citep{wild2006,sardane2015}. Further, the ratio $W_{\nai}/W_{\caii}\le$ 0.3 suggests that Ca is not as heavily depleted onto dust grains as expected in dense star forming regions \citep{welty1996}. 

The quasar sightline appears to pass through the outer optical disc of a foreground galaxy, i.e. $\sim$5$''$ north of the galaxy's centre, in the SDSS image \citep[see fig. 5 of][]{dutta2017a}. 
The redshift of this galaxy is obtained as 0.1 from South African Large Telescope (SALT) long-slit spectrum \citep[see Appendix C of][]{dutta2017a}. The host galaxy of the \mgii\ system could 
be located behind this foreground galaxy. However, as mentioned in Section~\ref{sec_sample}, nebular emission lines at the redshift of the \mgii\ absorption are not detected in the SDSS quasar 
spectrum. Further, no extended emission at the redshift of the \mgii\ absorption has been detected in the SALT two-dimensional long-slit spectrum of the quasar. We also identify a possible host 
galaxy ($z_{\rm phot}$ = 0.355 $\pm$ 0.088) at $\sim$21$''$ north-east of the quasar ($b\sim$100 kpc) in the SDSS images (Table~\ref{tab:sample}). Deep space-based observations would help to 
identify the true host galaxy of this system.
\subsection{\hi\ \21\ non-detections}
\label{sec_nondet}
The \taudvl\ limits estimated in case of the \hi\ \21\ non-detections suggest that these systems would be sub-DLAs, if we assume a spin temperature of 100 K and a \fc\ of unity (see Table~\ref{tab:radiopara}). 
Even if we assume that the gas is warmer, i.e. \ts\ = 1000 K, and \fc\ = 1, all except one of the systems with \hi\ \21\ non-detection would be sub-DLAs based on the \taudvl\ limits. Indeed, in case of the 
quasar J0209$-$0438, a sub-DLA with \nhi\ = 10$^{19}$ \cms\ is detected at the redshift of the \mgii\ system in the {\it Hubble Space Telescope $-$ Cosmic Origins Spectrograph (HST$-$COS)} UV spectrum 
\citep{muzahid2015}. Hence, we do not have sufficient sensitivity to detect absorption from cold ($\sim$100 K) \hi\ gas in this case. Additionally, we note that the background radio source is resolved in our 
GMRT image (see Fig.~\ref{fig:overlays}), as well as in the VCS 2.3 GHz sub-arcsecond-scale image. Using \fc\ estimated from this image (Table~\ref{tab:sample}), the \nhi\ measurement and \taudvl\ constraint 
(Table \ref{tab:radiopara}), we estimate \ts\ $>$ 37 K. Further, the non-detection of H$_2$ absorption from this sub-DLA ($N$(H$_2$) $<$ 10$^{14}$ \cms) by \citet{muzahid2015}, indicates that most of the 
gas is likely to be warm. \nhi\ measurements for all the systems in our sample would help to interpret the \hi\ \21\ non-detections. Lastly, the large impact parameters of the candidate host galaxies from the 
quasar sightlines (based on photometric redshifts; see Table~\ref{tab:sample}) could explain the lack of \hi\ \21\ absorption in most of the systems in our sample.
%
%=========================== DISCUSSION ====================================================================================
%
\section{Discussion}
\label{sec_discussion}
\subsection{Detection rate of \hi\ \21\ absorption} 
\label{sec_discussion1}
We can estimate the detection rate of \hi\ \21\ absorption (\c21) as the fraction of systems showing \hi\ \21\ detections with \taudvl\ $\le$ \t0\ and \taudv\ $\ge$ \t0, where \t0\ is a
3$\sigma$ optical depth sensitivity. We use \t0\ = 0.30 \kms\ to estimate \c21\ in this work, which corresponds to a sensitivity of \nhi\ $\le$ 5 $\times$ 10$^{19}$ \cms\ for \ts\ = 100 K 
and \fc\ = 1, and facilitates comparisons with \hi\ \21\ absorption studies in the literature. We obtain $C_{21}^{\rm Mg~\textsc{ii}}$ = 0.18$^{+0.24}_{-0.12}$ at $z\sim$ 0.35 for our 
sample of strong \mgii\ systems. The quoted errors represent Gaussian 1$\sigma$ confidence intervals computed using tables of \citet{gehrels1986} assuming a Poisson distribution. 
We caution that due to the small sample size, the detection rate presented here has large uncertainty. The only other systematic survey of \hi\ \21\ absorption in \mgii-selected 
systems at the redshift range of our interest is presented in \citet{lane2000}. There are four systems in the sample of \citet{lane2000} that satisfy our selection criteria (Section~\ref{sec_sample}) 
\footnote{Here we have not considered the \zabs\ = 0.4367 system towards 1243$-$072 \citep{lane2001,kanekar2002}. This was identified as a \mgii\ system by \citet{wright1979} based on low-resolution 
spectrum. However, there is no published spectra or metal equivalent width information for this system.}. Three of these systems (\zabs\ = 0.3939 towards 0248$+$430, \zabs\ = 0.3130 towards 
1127$-$145, \zabs\ = 0.3950 towards 1229$-$021) show \hi\ \21\ absorption. The remaining system (\zabs\ = 0.4246 towards 0735$+$178) is a non-detection, however its reported optical depth 
limit is larger than our adopted optical depth sensitivity of \t0\ = 0.30 \kms. Hence, combining our sample with the above systems from the sample of \citet{lane2000} leads to 
$C_{21}^{\rm Mg~\textsc{ii}}$ = 0.36$^{+0.24}_{-0.15}$ at $z\sim$ 0.35.

Next, from the detection rate of \hi\ \21\ absorption in strong \mgii\ systems and that of DLAs in strong \mgii\ systems, we can infer the detection rate of \hi\ \21\ absorption in DLAs.
Table 1 of \citet{rao2006} shows that the fraction of strong \mgii\ systems that are DLAs is 0.27$^{+0.08}_{-0.06}$ at $z<$ 1. If we assume that the incidence of DLAs in strong \mgii\ systems 
remains constant at $z<$ 1 and consider our sample of strong \mgii\ systems, then the inferred incidence of \hi\ \21\ absorption in $z\sim$ 0.35 DLAs is $C^{\rm DLA}_{21}$ = 0.67$^{+0.33}_{-0.47}$ 
[note that this becomes unity when we include the systems from \citet{lane2000}]. This is consistent with the estimates obtained from \hi\ \21\ absorption studies of $z<$ 1 DLAs 
\citep[see][and references therein]{dutta2017a}, and that inferred from 0.5 $<z<$ 1.0 strong \mgii\ and \feii\ systems \citep[][]{gupta2012,dutta2017b}. Hence, the probability of 
detecting cold \hi\ gas appears to be higher (by about four times) in DLAs compared to strong \mgii-selected systems.

From an absorption-blind survey of quasar-galaxy pairs (QGPs), \citet{dutta2017a} have found that the detection rate of \hi\ \21\ absorbers is 0.16$^{+0.07}_{-0.05}$ within $b\sim$ 35 kpc of 
$z<0.4$ galaxies. Further, the \mgii\ absorber-galaxy catalog of \citet{nielsen2013} shows that the detection rate of strong \mgii\ absorbers at 0.3 $<z<$ 0.5 is 0.42$^{+0.28}_{-0.18}$ when the host 
galaxies are within $b\sim$ 35 kpc of the quasar sightlines. Hence based on the above, we expect the detection rate of \hi\ \21\ absorption in strong \mgii\ systems to be 0.38$^{+0.30}_{-0.20}$ at $b\le$ 
35 kpc. This is consistent with the \hi\ 21\ detection rate in our sample (0.40$^{+0.53}_{-0.26}$ at $b\le$ 35 kpc), if we assume that the nearest galaxy based on photometric redshift is the host galaxy, 
and that the host galaxy in case of J1551$+$0713 is located behind the foreground galaxy superimposed on the quasar (see Table~\ref{tab:sample}). Hence, broadly we can conclude that $\sim$40\% of the 
\hi\ gas within $b\sim$ 35 kpc of low-$z$ galaxies will produce strong \mgii\ absorption, and $\sim$40\% of these strong \mgii\ absorbers will produce detectable \hi\ \21\ absorption. This picture is 
consistent with the detection rates of \hi\ \21\ absorption in QGPs and our sample of strong \mgii\ systems.
\subsection{Redshift evolution of \hi\ \21\ absorbers} 
\label{sec_discussion2}
\begin{figure*}
\includegraphics[height=0.35\textheight, angle=90]{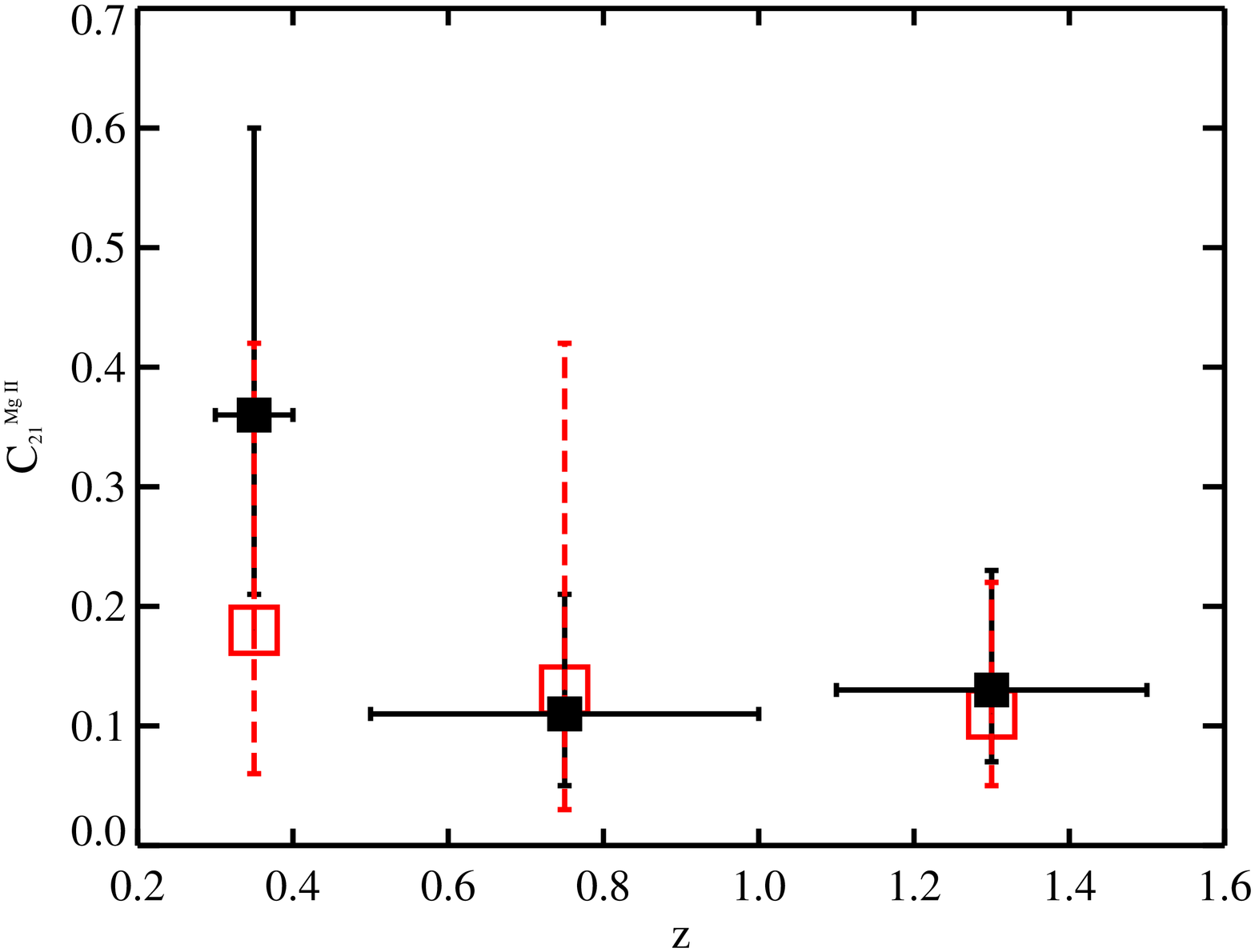}
\includegraphics[height=0.35\textheight, angle=90]{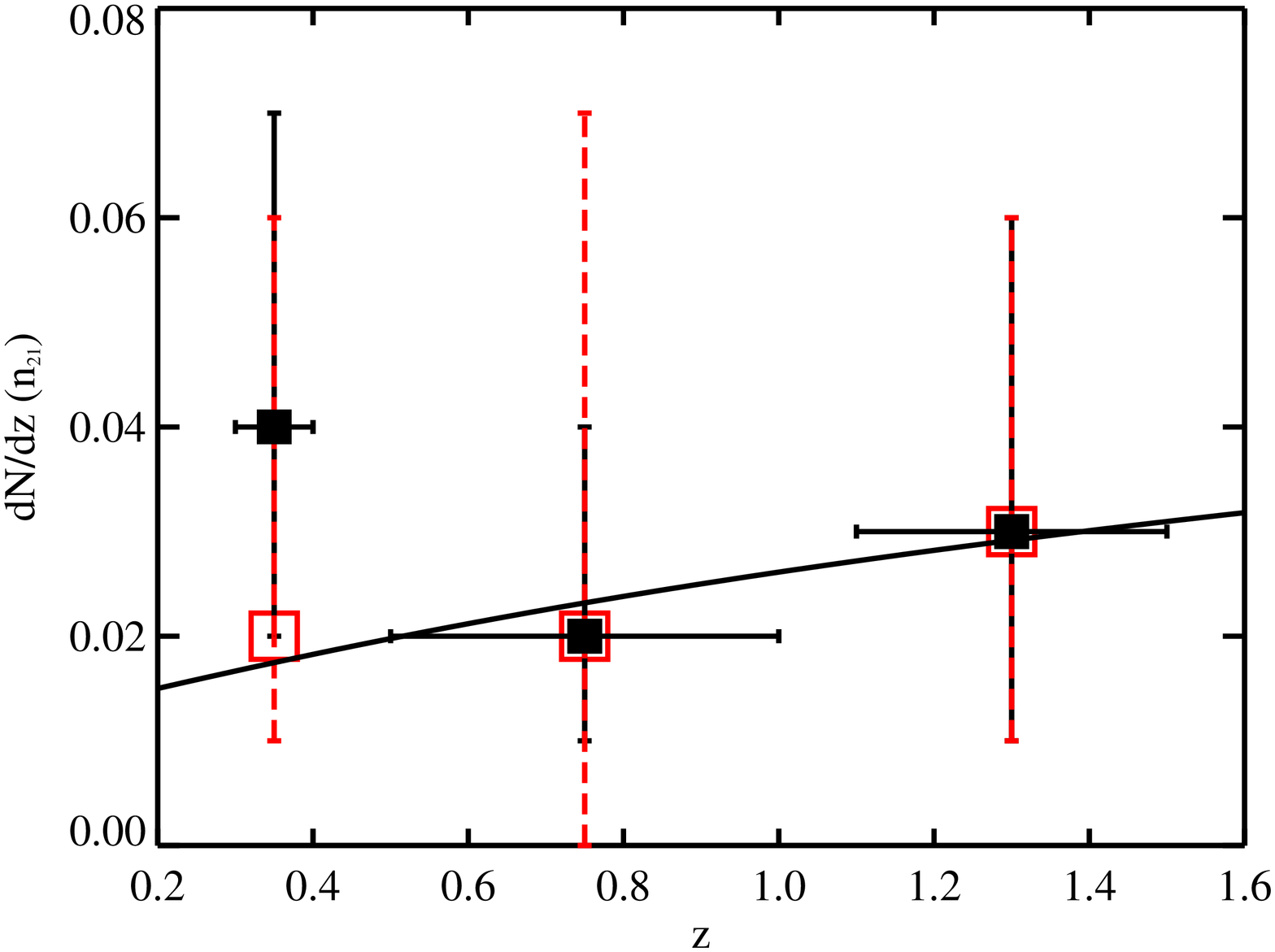}
\caption[Redshift evolution of incidence and number density per unit redshift of \hi\ \21\ absorbers in strong \mgii\ systems.]
{Redshift evolution of detection rate (for \t0\ = 0.3\,\kms; left) and number density per unit redshift (right) of \hi\ \21\ absorbers in strong \mgii\ systems. The open symbols (and dashed error bars) 
correspond to estimates based on our sample and that of \citet{gupta2009} and \citet{gupta2012}. The filled symbols correspond to estimates upon adding the systems from \citet{lane2000} and \citet{kanekar2009}. 
The solid line is the curve for non-evolving population of \hi\ \21\ absorbers normalized at $z$ = 1.3.}
\label{fig:evolution}
\end{figure*}

To study the redshift evolution of \hi\ \21\ absorbers in strong \mgii-selected systems, we consider our present sample and the samples of \citet{lane2000}, \citet{kanekar2009}, \citet{gupta2009}
and \citet{gupta2012}, for which both the \hi\ \21\ detections and non-detections have been systematically reported, since the inclusion of individually reported detections from the literature can bias 
the detection rate estimates. We compare $C_{21}^{\rm Mg~\textsc{ii}}$ (for \t0\ = 0.3\,\kms) over 0.3 $<z<$ 1.5 in the left panel of Fig.~\ref{fig:evolution}. We show in open symbols the detection rates 
considering only our sample and that of \citet{gupta2009} and \citet{gupta2012}, which are homogeneous in terms of sample selection, and treatment of optical and radio data. The detection rates upon adding 
the systems from \citet{lane2000} and \citet{kanekar2009} are shown in filled symbols. It can be seen that due to the large uncertainties, it is difficult to interpret the redshift evolution of 
$C_{21}^{\rm Mg~\textsc{ii}}$. Considering only our measurements, the incidence of \hi\ \21\ absorption in strong \mgii\ systems does not appear to be evolving significantly over 0.3 $<z<$ 1.5, within the 
uncertainties. If the measurements of \citet{lane2000} are considered, there is an indication that the incidence of \hi\ \21\ absorption in strong \mgii\ systems may have increased from 0.5 $<z<$ 1.5 to 
0.3 $<z<$ 0.5 by a factor of three, though the evolution is not statistically significant. Note that due to lack of low-frequency VLBA images for all the radio sources in our sample, we do not correct the 
measurements for \fc. The detection rates for different \mgii\ samples at 0.5 $<z<$ 1.5 when corrected for partial coverage are listed in table 5 of \citet{gupta2012}.

Next, as described in \citet{gupta2009,gupta2012}, we can estimate the number density per unit redshift of \hi\ \21\ absorbers ($n_{21}$) with \taudv\ $\ge$ \t0\ and \wmg\ $\ge$ 1.0 \AA, 
from $C_{21}^{\rm Mg~\textsc{ii}}$ and the number density per unit redshift of \mgii\ absorbers ($n_{\rm Mg~\textsc{ii}}$) with \wmg\ $\ge$ 1.0 \AA\ 
\footnote{$n_{\rm Mg~\textsc{ii}}$($W_0,z$) = $n_0\times(1 + z)^\gamma$, where $n_0$ = 0.080 $^{+0.015}_{-0.005}$ and $\gamma$ = 1.40 $\pm$ 0.16 for $W_0 >$ 1 \AA\ \citep{prochter2006}.}. 
We obtain $n_{21}$ = 0.02$^{+0.04}_{-0.01}$ at $z$ = 0.35 considering only our sample, and $n_{21}$ = 0.04$^{+0.03}_{-0.02}$ upon including the systems from \citet{lane2000} (see Section~\ref{sec_discussion1}).
This is consistent, within the uncertainties, with the estimate of $n_{21}$ at $z$ = 0.1 obtained from the detection rate of \hi\ \21\ absorption in low-$z$ QGPs, assuming a characteristic 
radius of 35 kpc \citep[see][]{dutta2017a}. Similar to $C_{21}^{\rm Mg~\textsc{ii}}$, within the large uncertainties we observe no significant evolution of $n_{21}$ over 0.3 $<z<$ 1.5 (see 
right panel of Fig.~\ref{fig:evolution}). The $n_{21}$ estimates are also consistent with a non-evolving population (i.e. no intrinsic redshift evolution in physical parameters for the assumed 
cosmology) of \hi\ \21\ absorbers. \\
\section{Summary}
\label{sec_summary}
We have presented here the results from our GMRT search for \hi\ \21\ absorption in eleven strong \mgii\ systems at 0.3 $<z<$ 0.5 selected from SDSS. The main results from our study are the following:
\begin{enumerate}
 \item We report two new \hi\ \21\ absorption detections. The FWHM of the absorption lines at \zabs\ = 0.3940 towards J1428$+$2103 and at \zabs\ = 0.3289 towards J1551$+$0713, indicate that the gas 
 kinetic temperature (and hence the spin temperature) is $<$1600 K and $<$350 K, respectively. The small velocity widths are also consistent with the correlation found between $v_{90}$ and \zabs\ by 
 \citet{dutta2017b}.
 \item The lack of \hi\ \21\ detection in the other nine \mgii\ systems could be because they are arising from sub-DLAs, from which we do not have sufficient optical depth sensitivity to detect cold 
 $\sim$100 K gas. This is indeed the case for the system towards J0209$-$0438, where \nhi\ measurement from {\it HST} spectra is available. 
 \item The non-detection of \hi\ \21\ absorption is also consistent with most of the systems having candidate  host galaxies with $b>$ 35 kpc, where there has been no detection of \hi\ \21\ absorption 
 around $z<$ 0.4 galaxies \citep[see][]{curran2016,dutta2017a}. This reinforces the argument that just \wmg\ $\ge$ 1 \AA\ selects a wide range of impact parameters, and hence additional constraints like 
 \wfe\ $\ge$ 1 \AA\ are required to probe high \nhi\ cold gas at small impact parameters from galaxies \citep{dutta2017b}. UV spectra covering Lyman-$\alpha$ and \feii\ absorption from the \mgii\ systems 
 would help to interpret the incidence of \hi\ \21\ absorption at these redshifts.
 \item By comparing with \hi\ \21\ studies in samples of 0.5 $<z<$ 1.5 strong \mgii\ systems from the literature \citep[as compiled in][]{gupta2012}, we do not find any significant evolution in the incidence 
 and number density per unit redshift of \hi\ \21\ absorbers in strong \mgii\ systems over 0.3 $<z<$ 1.5, within the uncertainties. Upcoming blind \hi\ \21\ surveys with the Square Kilometre Array pre-cursors 
 are expected to provide  accurate and uniform measurement of the redshift evolution of the cold gas fraction in galaxies upto $z\sim$ 1.5. \\
\end{enumerate}
%
%========================================= Acknowledgment ==================================================================
%
%\newline \newline
\noindent \textbf{ACKNOWLEDGEMENTS} 
\newline 
\noindent 
We thank the referee, Dr. Wendy Lane, for her useful comments.
We thank the staff at GMRT for their help during the observations. 
GMRT is run by the National Centre for Radio Astrophysics of the Tata Institute of Fundamental Research. 
RS and NG acknowledge the support from Indo-Fench centre for the promotion of Advanced Research (IFCPAR) under Project No. 5504-2.
This research has made use of the NASA/IPAC Extragalactic Database (NED) which is operated by the Jet Propulsion Laboratory, California Institute of Technology, 
under contract with the National Aeronautics and Space Administration. 

Funding for SDSS-III has been provided by the Alfred P. Sloan Foundation, the Participating Institutions, the National Science Foundation, and the U.S. Department of Energy Office of Science. 
The SDSS-III web site is http://www.sdss3.org/. SDSS-III is managed by the Astrophysical Research Consortium for the Participating Institutions of the SDSS-III Collaboration including the 
University of Arizona, the Brazilian Participation Group, Brookhaven National Laboratory, Carnegie Mellon University, University of Florida, the French Participation Group, the German Participation 
Group, Harvard University, the Instituto de Astrofisica de Canarias, the Michigan State/Notre Dame/JINA Participation Group, Johns Hopkins University, Lawrence Berkeley National Laboratory, Max Planck 
Institute for Astrophysics, Max Planck Institute for Extraterrestrial Physics, New Mexico State University, New York University, Ohio State University, Pennsylvania State University, University of 
Portsmouth, Princeton University, the Spanish Participation Group, University of Tokyo, University of Utah, Vanderbilt University, University of Virginia, University of Washington, and Yale University. 
%
%======================================== Bibliography =====================================================================
% 
\def\aj{AJ}%
\def\actaa{Acta Astron.}%
\def\araa{ARA\&A}%
\def\apj{ApJ}%
\def\apjl{ApJ}%
\def\apjs{ApJS}%
\def\ao{Appl.~Opt.}%
\def\apss{Ap\&SS}%
\def\aap{A\&A}%
\def\aapr{A\&A~Rev.}%
\def\aaps{A\&AS}%
\def\azh{A$Z$h}%
\def\baas{BAAS}%
\def\bac{Bull. astr. Inst. Czechosl.}%
\def\caa{Chinese Astron. Astrophys.}%
\def\cjaa{Chinese J. Astron. Astrophys.}%
\def\icarus{Icarus}%
\def\jcap{J. Cosmology Astropart. Phys.}%
\def\jrasc{JRASC}%
\def\mnras{MNRAS}%
\def\memras{MmRAS}%
\def\na{New A}%
\def\nar{New A Rev.}%
\def\pasa{PASA}%
\def\pra{Phys.~Rev.~A}%
\def\prb{Phys.~Rev.~B}%
\def\prc{Phys.~Rev.~C}%
\def\prd{Phys.~Rev.~D}%
\def\pre{Phys.~Rev.~E}%
\def\prl{Phys.~Rev.~Lett.}%
\def\pasp{PASP}%
\def\pasj{PASJ}%
\def\qjras{QJRAS}%
\def\rmxaa{Rev. Mexicana Astron. Astrofis.}%
\def\skytel{S\&T}%
\def\solphys{Sol.~Phys.}%
\def\sovast{Soviet~Ast.}%
\def\ssr{Space~Sci.~Rev.}%
\def\zap{$Z$Ap}%
\def\nat{Nature}%
\def\iaucirc{IAU~Circ.}%
\def\aplett{Astrophys.~Lett.}%
\def\apspr{Astrophys.~Space~Phys.~Res.}%
\def\bain{Bull.~Astron.~Inst.~Netherlands}%
\def\fcp{Fund.~Cosmic~Phys.}%
\def\gca{Geochim.~Cosmochim.~Acta}%
\def\grl{Geophys.~Res.~Lett.}%
\def\jcp{J.~Chem.~Phys.}%
\def\jgr{J.~Geophys.~Res.}%
\def\jqsrt{J.~Quant.~Spec.~Radiat.~Transf.}%
\def\memsai{Mem.~Soc.~Astron.~Italiana}%
\def\nphysa{Nucl.~Phys.~A}%
\def\physrep{Phys.~Rep.}%
\def\physscr{Phys.~Scr}%
\def\planss{Planet.~Space~Sci.}%
\def\procspie{Proc.~SPIE}%
\let\astap=\aap
\let\apjlett=\apjl
\let\apjsupp=\apjs
\let\applopt=\ao
\bibliographystyle{mnras}
\bibliography{mybib}
\bsp
\label{lastpage}
\end{document}